\documentclass[12pt]{iopart}
\usepackage{iopams}  
\usepackage{graphicx}
\begin{document}

\title[Tunnelling anisotropic magnetoresistance of Fe/GaAs/Ag]%
      {Tunnelling anisotropic magnetoresistance of Fe/GaAs/Ag(001)
       junctions from first principles: Effect of hybridized
       interface resonances}

\author{R S\'ykora$^1$ and I Turek$^2$}

\address{$^1$Charles University, Faculty of Mathematics and Physics,
               Department of Condensed Matter Physics,
               Ke Karlovu 5, CZ-12116 Prague, Czech Republic}

\address{$^2$Institute of Physics of Materials, Academy of Sciences of
the Czech Republic, \v{Z}i\v{z}kova~22, CZ-61662 Brno, Czech Republic}

\eads{\mailto{rudolf.sykora@gmail.com}, \mailto{turek@ipm.cz}} 

\begin{abstract}
Results of first-principles calculations of the Fe/GaAs/Ag(001)
epitaxial tunnel junctions reveal that hybridization of interface
resonances formed at both interfaces can enhance the tunnelling
anisotropic magnetoresistance (TAMR) of the systems.
This mechanism is manifested by a non-monotonic dependence of the
TAMR effect on the thickness of the tunnel barrier, with a maximum
for intermediate thicknesses.
A detailed scan of $\bi{k}_\|$-resolved transmissions over
the two-dimensional Brillouin zone proves an interplay between a few
hybridization-induced hot spots and a contribution to the tunnelling
from the vicinity of the $\bar{\Gamma}$ point.
This interpretation is supported by calculated properties of
a simple tight-binding model of the junction which reproduce
qualitatively most of the features of the first-principles theory.
\end{abstract}


\section{Introduction\label{s_intr}}

Systems and devices with applicability in spintronics include
traditional magnetic multilayers and tunnel junctions consisting of
several magnetic layers \cite{r_2008_smt, r_2010_bp, r_2012_tz}
as well as more recent artificial structures containing only a single
magnetic part \cite{r_2010_mga, r_2011_pwm}.
The former exhibit the well-known giant and tunnelling magnetoresistance
effects arising due to changes in the mutual orientation of
magnetization directions, the latter are featured especially by
properties driven by the spin-orbit (SO) interaction.
The so-called tunnelling anisotropic magnetoresistance (TAMR) was
observed in systems FM/I/NM, where the FM denotes a ferromagnetic
metallic electrode, the NM denotes a non-magnetic metallic electrode,
and the I denotes a non-magnetic insulating (semiconducting)
barrier \cite{r_2004_grj, r_2007_mms}.

The TAMR effect observed in conventional Fe/GaAs/Au tunnel
junctions \cite{r_2007_mms} proved to be rather weak which
prevents its direct use.
Similarly, a complementary phenomenon, namely, the current
(or voltage) induced spin-transfer torques \cite{r_2010_mga},
which might be employed for magnetization switching in the
FM/I/NM devices, requires a large TAMR effect as a necessary
prerequisite for efficient spintronic devices \cite{r_2011_am}.
Recent attempts to enhance the TAMR values included, e.g.,
modification of the surface structure of the semiconductor
layer during the growth process \cite{r_2011_uha} or the use
of an antiferromagnetic metal instead of the ferromagnetic one
in the magnetic electrode \cite{r_2011_pwm}.
The latter approach yields strongly enhanced TAMR values, which,
however, could be observed only at low temperatures, despite the
much higher N\'eel temperature of the antiferromagnetic material.

The TAMR attracted interest also on theoretical side.
A number of various topics were addressed in the framework
of phenomenological models.
These approaches discussed the role of the anisotropic density
of states of the FM electrode \cite{r_2004_grj},
the symmetry properties and the interplay of the Rashba
and Dresselhaus contributions to the SO interaction,
the effect of external magnetic fields and applied bias
voltages \cite{r_2007_mms, r_2009_mf}, etc. 
\emph{Ab initio} calculations of the TAMR were carried out
for tunnel junctions with Fe and Cu electrodes separated
by vacuum \cite{r_2007_cbt} and GaAs \cite{r_2007_cbs}
as well as for the Fe/GaAs/Au trilayers \cite{r_2010_ebb}.
The first-principles studies proved that interface states
(resonances) formed at the FM/I interface play an important
role in the TAMR phenomenon.

All existing theoretical studies ascribe the TAMR primarily to
electronic properties of the FM electrode, the tunnel barrier and
their interface, whereas the NM electrode and the I/NM interface
are considered to be of secondary importance.
In the present paper, we show by means of first-principles
calculations for the Fe/GaAs/Ag system and by the study of
a simple tight-binding (TB) model Hamiltonian, that the standard
picture of the TAMR is not generally valid.
In particular, we find that interface resonances at the I/NM
interface--when hybridized with those at the FM/I interface--can
yield high TAMR values.
Moreover, we predict a non-monotonic dependence of the TAMR value
on the thickness of the tunnelling barrier and discuss its physical
origin.

The paper is organized as follows.
The structural model of the Fe/GaAs/Ag junction and the
\emph{ab initio} techniques employed are summarized
in section~\ref{s_meth} while the obtained results and their
discussion are presented in section~\ref{s_resd}:
the properties of the GaAs/Ag interface are presented in
section~\ref{ss_esag} and those of the Fe/GaAs/Ag system
are contained in section~\ref{ss_cond}.
The TB model is formulated and studied in section~\ref{s_hirtbm}
and the summary of the main results is given in the last section.

\section{Model and methods\label{s_meth}}

The structural model of the Fe/GaAs/Ag(001) tunnel junction
represents a simple generalization of the model for the
Fe/GaAs/Cu trilayers \cite{r_2007_cbs}.
The slab of the zinc-blende (zb) structure of GaAs is attached
epitaxially to semiinfinite leads of the body-centered cubic (bcc)
Fe and the face-centered cubic (fcc) Ag; the atomic planes are
the (001) planes of all three structures.
The structure of the Fe/GaAs interface has been assumed without
any reconstructions and layer relaxations on the basis of the 
ideal ratio of the lattice parameters 
$a_\mathrm{zb}/a_\mathrm{bcc} = 2$, that is satisfied with a good
accuracy for GaAs and Fe.
The structure of the GaAs/Ag interface was also assumed without
reconstructions; it employed the ideal ratio 
$a_\mathrm{zb}/a_\mathrm{fcc} = \sqrt{2}$ and the mutual rotation
of both bulk structures by $\pi/4$ around the common (001) axis.
The interplanar distance between the adjacent atomic planes of
the GaAs and Ag parts was set equal to the arithmetic average of
the distances in the bulk GaAs ($a_\mathrm{zb}/4$) and the bulk
Ag ($a_\mathrm{fcc}/2$). 
The $z$ axis of the coordinate system is perpendicular to
the atomic planes, 
while the $x$ and $y$ axes coincide respectively with the [100] 
and [010] directions of the fcc lattice, i.e., they point along
the $[1, \pm 1, 0]$ directions of the bcc and zb lattices.
The GaAs slab contains $n$ As-atomic (001) planes with As-termination
on both sides; the investigated systems are thus
abbreviated as Fe/As(GaAs)$_{n-1}$/Ag(001).

The electronic structure of the systems was calculated by
self-consistent scalar relativistic tight-binding linear
muffin-tin orbital (TB-LMTO) 
method \cite{r_1984_aj, r_1997_tdk, r_2000_tkd}.
The calculations were performed using the local spin-density
approximation with the Vosko-Wilk-Nusair parametrization
of the exchange-correlation potential \cite{r_1980_vwn}.
The LMTO valence basis $Ls$, where $L = (\ell, m)$ is the orbital
index and $s = \uparrow, \downarrow$ is the spin index, was limited
to $\ell \le 2$ and the so-called empty spheres were used for an
efficient treatment of the open zb structure.
The low-lying Ga-$3d$ orbitals were treated as core orbitals
so that the valence basis comprised the Ga-$4d$ 
orbitals \cite{r_2006_tck}.
This choice leads to a good description of the bulk bandstructure
of the GaAs concerning both the band gap (around 1.2 eV) and the
valence bandwidth (around 6.8 eV), which agree quite well with
measured data \cite{r_1980_cka}.

The study of the TAMR was based on the conductances evaluated in the
current-perpendicular-to-the-planes (CPP) geometry.
For this purpose, the scalar relativistic TB-LMTO Hamiltonian was
completed by adding an on-site SO term having a simple 
$\xi \bi{L} \cdot \bi{S}$ form, where the atomic-like
SO parameters $\xi_{\bi{R}\ell,ss'}$ of the atom
at the lattice site $\bi{R}$ were calculated
from the self-consistent electronic structure \cite{r_2008_tdk}.
The matrix elements of the SO term have the form
\begin{equation}
H^\mathrm{SO}_{\bi{R}Ls,\bi{R}'L's'} =
\delta_{\bi{R}\bi{R}'} \delta_{LL'} 
\xi_{\bi{R}\ell,ss'} \langle L s | 
\bi{L} \cdot U^+ \bi{S} U | L' s' \rangle ,
\label{eq_meso}
\end{equation}
where the symbols $\bi{L}$ and $\bi{S}$ denote respectively
the orbital and spin angular momentum operators. 
The direction of the FM electrode magnetization is given by two
angles $\theta$ and $\phi$, or by the unit vector $\bi{n} = 
(\sin\theta \cos\phi, \sin\theta \sin\phi, \cos\theta)$. 
These quantities enter the Hamiltonian via the operator $U$ in
$H^\mathrm{SO}$ (\ref{eq_meso}), which acts only on the
spin indices $s, s'$ and which is represented by the 
spin-1/2 rotation matrix $U=D^{(1/2)}(\phi,\theta,0)=
\exp(-\mathrm{i}\phi\sigma_z/2) \exp(-\mathrm{i}\theta\sigma_y/2)$,
where the $\sigma_y$ and $\sigma_z$ are two of the Pauli spin
matrices \cite{r_1961_mer}.
Note that the form of (\ref{eq_meso}) corresponds to the global
rotation of the spin quantization axis, whereas the coordinate
system for the orbital motion remains unchanged.

The CPP conductances were evaluated in the TB-LMTO Kubo-Landauer
formalism \cite{r_2000_kdb, r_2006_ctk} as averages of the
$\bi{k}_\|$-dependent transmissions $T(\bi{k}_\|)$ over
the two-dimensional (2D) Brillouin zone (BZ). 
The mixing of both spin channels due to the SO interaction has
been implemented similarly to the case of non-collinear layered 
spin structures \cite{r_2000_kdb, r_2009_ct}.
For thicknesses of the GaAs barriers relevant in experiments
($n > 20$), the tunnelling current is carried mainly by the
states with $\bi{k}_\|$-vectors from a small central region of
the whole 2D BZ. 
A sufficiently dense mesh of sampling points (corresponding to 
$5.4 \times 10^5$ $\bi{k}_\|$-vectors in the full 2D BZ) has
been used to get reliable values of the CPP conductances. 

\section{Results and discussion\label{s_resd}}

\subsection{Electronic structure of Ag/GaAs/Ag(001)
            systems\label{ss_esag}}

In recent theoretical studies, the Fe/GaAs(001) system
has been attached to a hypothetical bcc(001) 
Cu electrode since the bulk bcc Cu has a free-electron-like 
bandstructure and the GaAs/Cu interface has a featureless
transmission function \cite{r_2007_cbt, r_2007_cbs}.
This setup is advantageous for investigations of the role of the
Fe/GaAs(001) interface state (resonance) lying in the minority-spin
channel at the Fermi energy.
This interface resonance gives rise, e.g., to a reversal of spin 
polarization of the tunnelling current with applied 
voltage \cite{r_2007_cbs} and it can lead to a pronounced TAMR 
effect \cite{r_2007_cbt}.
Motivated by these facts, we have focused on electronic properties 
of the non-magnetic GaAs/Ag(001) interface prior to the study of the
Fe/GaAs/Ag junctions.

\begin{figure}
\begin{center}
\includegraphics[width=0.85\textwidth]{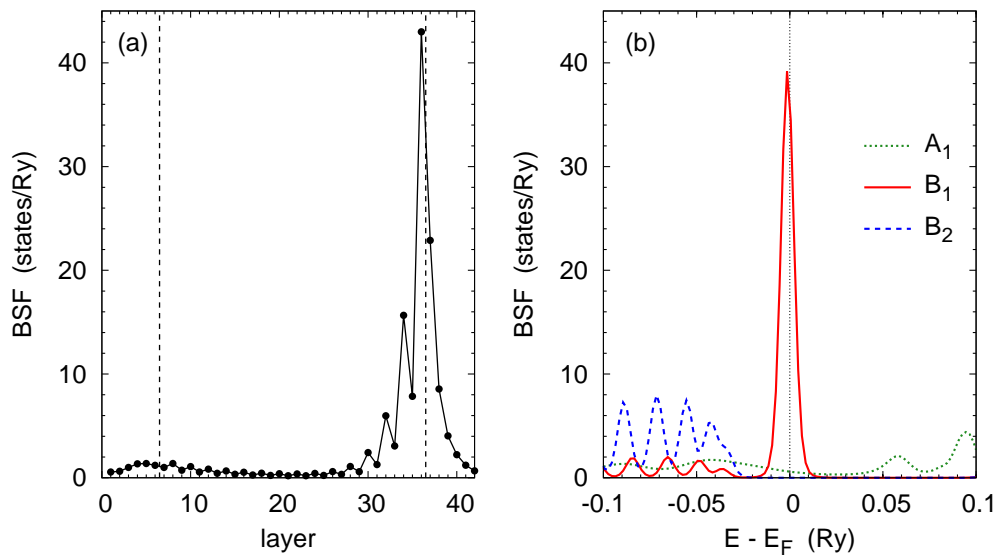}
\end{center}
\caption{%
The Bloch spectral functions (BSF) of the system 
Ag/(GaAs)$_{15}$/Ag(001) for $\bi{k}_\| = \bar{\Gamma}$:
(a) The layer-resolved BSF for the energy $E$ located 1.5 mRy
below the Fermi level. The dashed vertical lines denote the
left (Ag/Ga) and right (As/Ag) interfaces.
(b) The BSF's of the As interface atom as functions of energy,
resolved according to the irreducible representations of the
point group $C_{\rm 2v}$. The vertical line denotes the position
of the Fermi level of fcc Ag.
\label{f_bsf}}
\end{figure}

We have thus studied systems Ag/GaAs/Ag(001) with different
thicknesses and both terminations (Ga or As) of the GaAs barrier.
These studies were done first without the SO interaction. 
Since the tunnelling current is mostly carried by states with
$\bi{k}_\|$-vectors in the vicinity of the $\bar{\Gamma}$ point,
we paid special attention to $\bi{k}_\| = \bar{\Gamma}$.
Figure~\ref{f_bsf}(a) displays the layer-resolved Bloch spectral
function of the system Ag/(GaAs)$_{15}$/Ag(001) for an energy
slightly below the Fermi energy ($E - E_{\rm F} = -1.5$~mRy). 
One can see a clear indication of an interface state at the GaAs/Ag
interface, i.e., at the As-terminated GaAs barrier.
The amplitude of the interface state is maximal in the As layer
adjacent to the Ag electrode.
A similar interface state was found in the Au/GaAs/Au(001) system,
located however 10 mRy below the Fermi energy, whereas in the
Cu/GaAs/Cu(001) with bcc Cu, no such state appears, in agreement
with previous studies \cite{r_2007_cbs}.
It should be noted that no similar interface state was found at the
Ga-terminated boundary of the GaAs barrier in a wide energy interval
around the Fermi energy (inside the band gap of GaAs), irrespective
of the electrode metal (Cu, Ag, Au).

The origin of the interface state can be understood from the Bloch
spectral functions of the boundary As site resolved with respect to
the symmetry given by the point group of the interface, namely,
the $C_{\rm 2v}$ group.
This group has four one-dimensional irreducible representations:
A$_1$, A$_2$, B$_1$, and B$_2$ \cite{r_1960_vh},
of which only the A$_1$ (subspace spanned by orbitals $s$, $p_z$, 
$d_{z^2}$ and $d_{x^2-y^2}$) is compatible with the symmetry of 
propagating states of the Ag(001) electrode at the Fermi level. 
As can be seen in figure~\ref{f_bsf}(b), the interface state is
entirely of the symmetry B$_1$ (subspace spanned by orbitals $p_y$
and $d_{yz}$), which is incompatible with the propagating states
available in the Ag electrode.
This incompatibility is an important factor, since the semiinfinite
metallic electrode acts essentially like a vacuum half-space in
the formation of the interface state. 

\begin{figure}
\begin{center}
\rotatebox{270}{\includegraphics[width=0.42\textwidth]{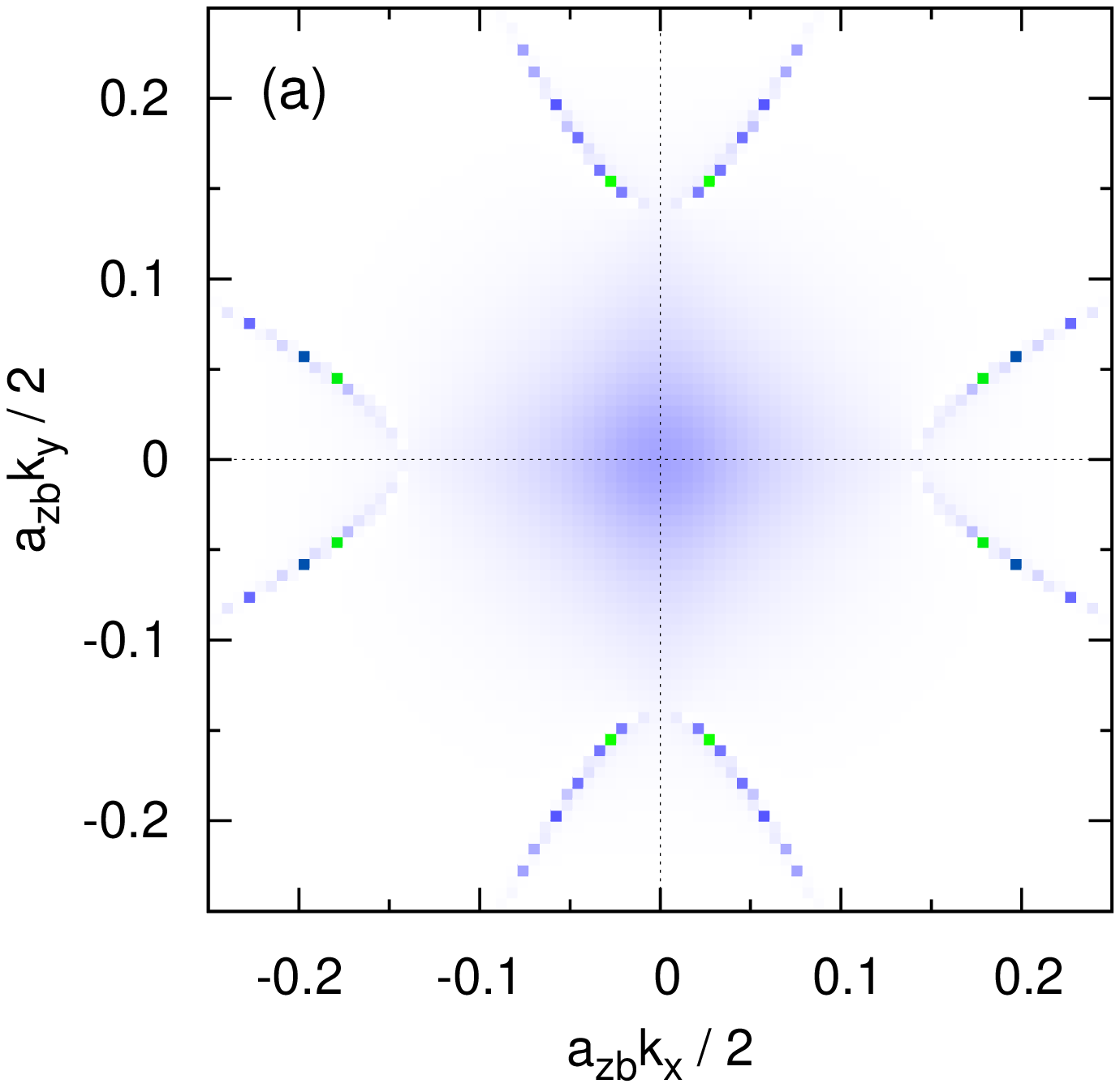}}
\rotatebox{270}{\includegraphics[width=0.42\textwidth]{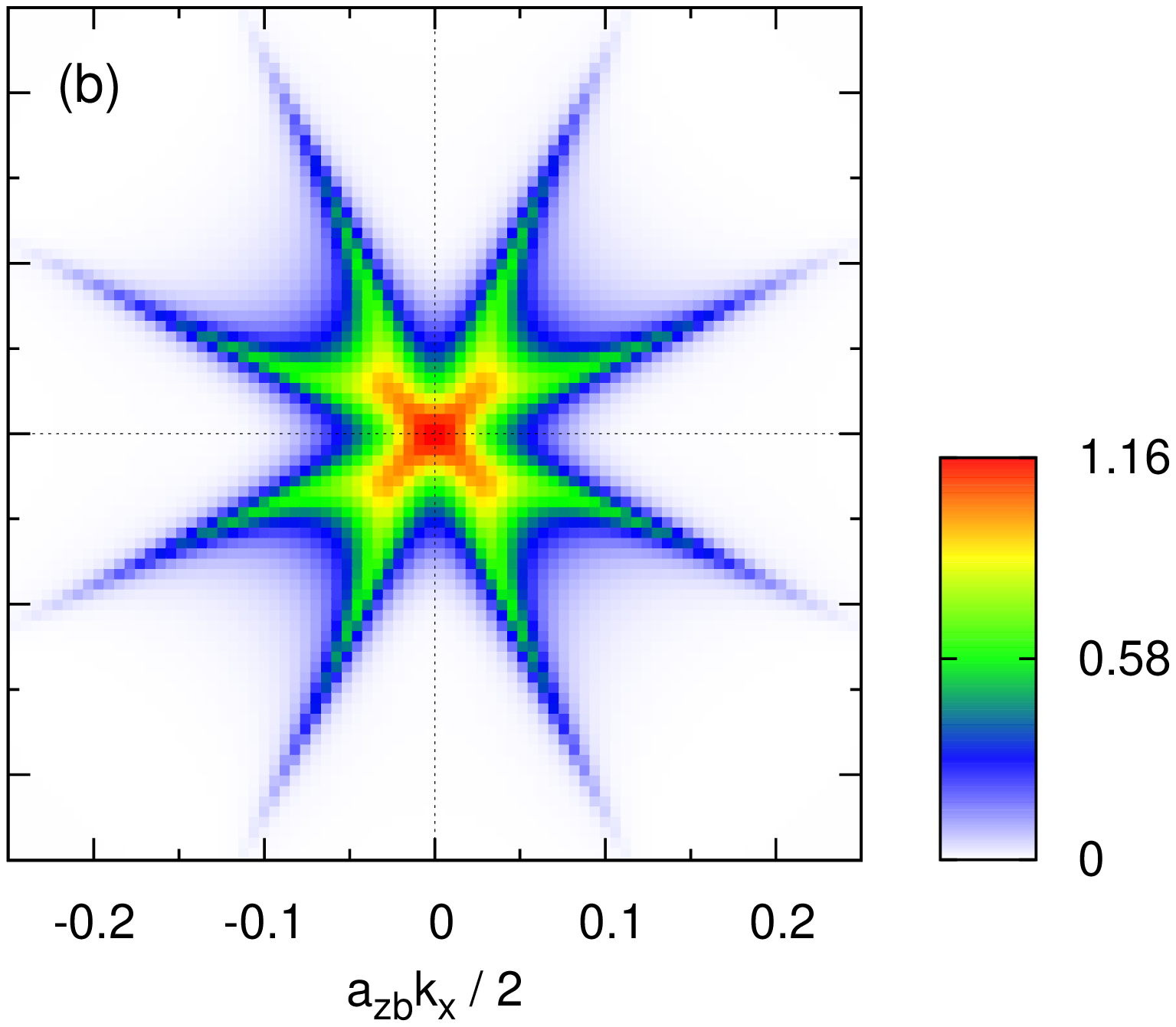}}
\end{center}
\caption{%
The $\bi{k}_\|$-resolved transmission of the
Ag/As(GaAs)$_{14}$/Ag(001) system at the Fermi energy: 
(a) without the SO interaction and (b) with the SO interaction.
The coloured scale of the $T(\bi{k}_\|)$ for both panels is 
shown on the right.
\label{f_kres_nm}}
\end{figure}

In the Landauer picture of the ballistic transport, interface
states do not contribute directly to the system conductance, since
the latter is given solely by the transmission coefficients between
the propagating channels of the leads \cite{r_1995_sd}.
However, if an interface state is coupled, e.g., by a weak
interaction to the propagating states, it can become a resonance
with a non-negligible effect on the conductance.
In the present case, the SO interaction provides such a coupling of
the B$_1$-like interface state to the A$_1$-like propagating state,
which follows from an analysis of the double group $C_{\rm 2v}$ and
its irreducible representations: all spin-orbitals belong to the
single additional two-dimensional representation of this
double group \cite{r_1957_gfk}.
The influence of the interface resonances on tunnelling is especially
strong in symmetric junctions with identical electrodes, where 
both resonances become hybridized across the tunnel barrier which
enhances the conductance appreciably \cite{r_2002_wpz}.
This phenomenon is illustrated in figure~\ref{f_kres_nm}, where the
$\bi{k}_\|$-resolved transmissions $T(\bi{k}_\|)$ are compared for
the symmetric junction Ag/As(GaAs)$_{14}$/Ag(001) treated without
and with SO coupling.
The pronounced enhancement of the $T(\bi{k}_\|)$ in vicinity of the
$\bar{\Gamma}$ point is clearly visible; the total conductance of
the junction increases by one order of magnitude due to the SO
interaction.
This result indicates importance of the GaAs/Ag interface for
the transport behaviour of the Fe/GaAs/Ag system.

\subsection{Conductances and TAMR of Fe/GaAs/Ag
            junctions \label{ss_cond}}

\begin{figure}
\begin{center}
\includegraphics[width=0.50\textwidth]{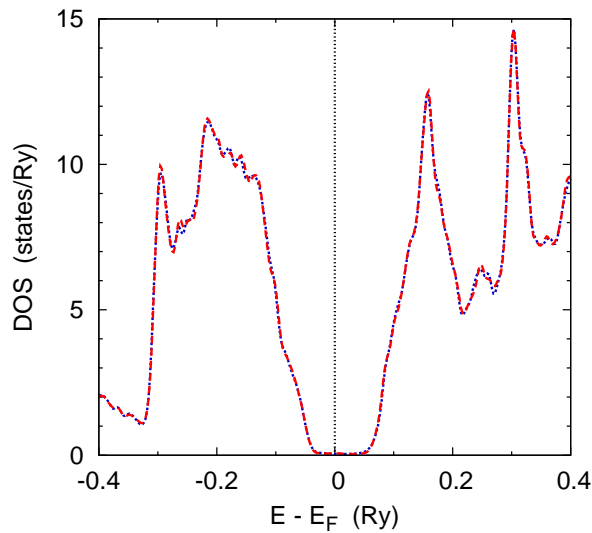}
\end{center}
\caption{%
The density of states in two neighbouring Ga- and As-atomic planes
in the middle of the Fe/As(GaAs)$_{10}$/Ag(001) system for the 
majority spin ($\dotted$, blue) and the minority spin 
($\dashed$, red).
The vertical line denotes the position of the Fermi level.
\label{f_dos}}
\end{figure}

\begin{figure}
\begin{center}
\includegraphics[width=0.80\textwidth]{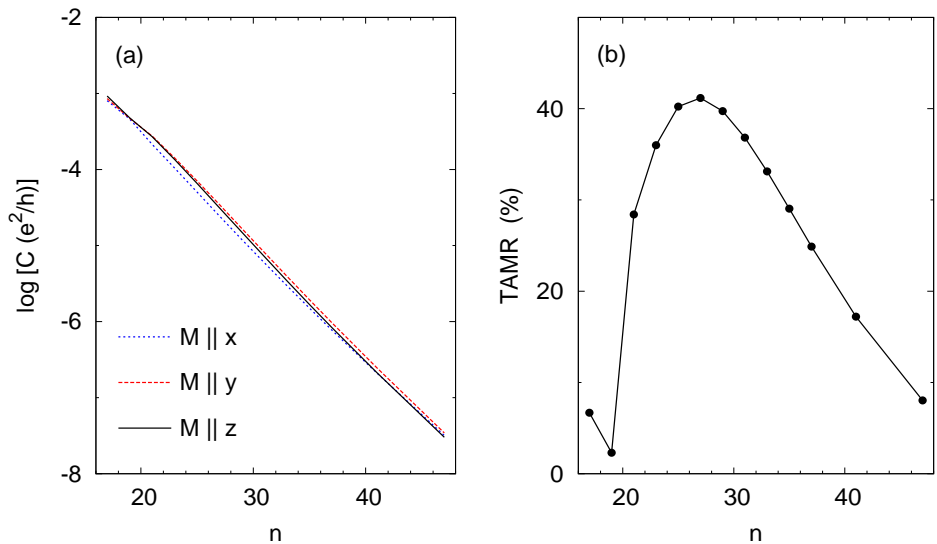}
\end{center}
\caption{%
Dependence of the transport properties of the tunnel junctions
Fe/As(GaAs)$_{n-1}$/Ag(001) on the thickness $n$:
(a) the conductances for the Fe magnetization pointing along the
$x$, $y$ and $z$ axis, and (b) the corresponding in-plane TAMR.
\label{f_thick}}
\end{figure}

\begin{figure}
\begin{center}
\includegraphics[width=0.60\textwidth]{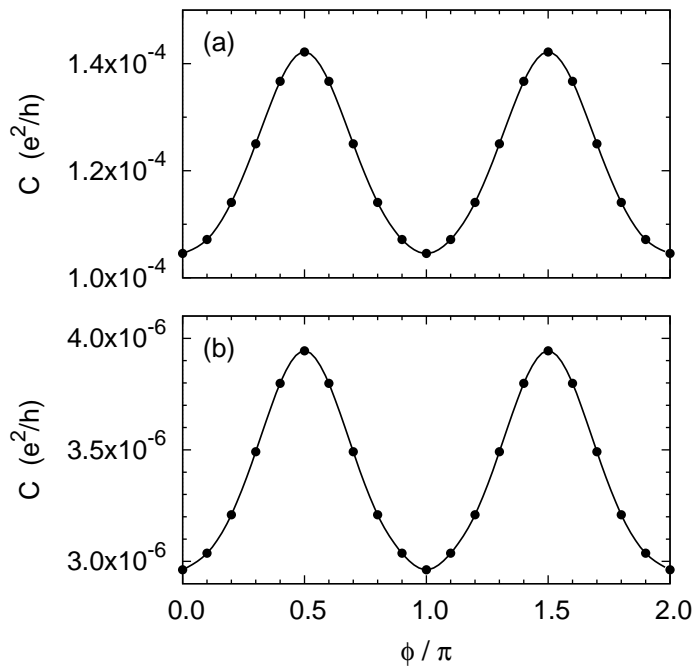}
\end{center}
\caption{%
The conductance $C(\pi/2, \phi)$ of the systems
Fe/As(GaAs)$_{n-1}$/Ag(001) as a function of the angle $\phi$:
(a) for $n=23$ and (b) for $n=33$.
\label{f_angle}}
\end{figure}

Figure~\ref{f_dos} shows the local density of states (DOS) of
the central GaAs layer (two neighbouring atomic planes)
inside the Fe/As(GaAs)$_{10}$/Ag(001) junction.
The shape of the DOS is bulk-like, with the band gap clearly formed
around the Fermi energy and with negligible spin polarization.
These features prove that the junction is in a tunnelling regime with
metal-induced gap states significantly suppressed for this and higher
barrier thicknesses.
The tunnelling regime is also manifested by an exponential decay of
the conductance with the increasing GaAs thickness $n$, plotted in
figure~\ref{f_thick}(a) for three orientations of the iron magnetic
moment, i.e., along the $x$, $y$, and $z$ axis.
For a given $n$, the conductance is obviously sensitive to the
magnetization direction which leads to the TAMR effect.
The orientational dependence was studied in detail for magnetization
directions in the atomic planes, i.e., as functions of the
angle $\phi$ with the fixed value of $\theta = \pi/2$.
The resulting angular dependences, shown in figure~\ref{f_angle} for
$n=23$ and $n=33$, reflect the two-fold rotation symmetry (point
group $C_{\rm 2v}$) of the system, in full agreement with previous
calculations \cite{r_2010_ebb} and measurements \cite{r_2007_mms}
performed for similar Fe/GaAs/Au(001) junctions.

The in-plane TAMR, defined from the angular dependence of the
conductance $C(\theta, \phi)$ as 
${\rm TAMR} = [C(\pi/2, \pi/2) - C(\pi/2, 0)]/C(\pi/2, 0)$, is
presented in figure~\ref{f_thick}(b) as a function of the barrier
thickness $n$.
The calculated TAMR effect is quite large, exceeding 10\%, in
reasonable agreement with the values calculated for the Fe/GaAs/Au
system \cite{r_2010_ebb}, but about two orders of magnitude stronger
than the experimentally observed TAMR values \cite{r_2007_mms}.
Moreover, the calculated thickness dependence is non-monotonic with
a maximum obtained around $n = 27$ corresponding to the GaAs 
thickness of about 7.5 nm.
In order to identify possible reasons for the calculated high TAMR
values and the non-monotonic thickness dependence, additional
analysis is needed.

\begin{table}
\caption{\label{tabone}
Dependence of the in-plane TAMR on the position of two atomic planes
with disorder in the Fe/As(GaAs)$_{26}$/Ag(001) junction.
The first row corresponds to the ideal system.}

\begin{indented}
\item[]
\begin{tabular}{lc}
\br
position & TAMR (\%) \cr
\mr
--- &  41.2 \cr
inside Fe & 39.4 \cr
at Fe/As & \phantom{0}3.7 \cr
inside GaAs & 41.4 \cr
at As/Ag & \phantom{0}4.7 \cr
inside Ag &  32.2 \cr
at Fe/As and As/Ag & \phantom{0}3.6 \cr
\br
\end{tabular}
\end{indented}
\end{table}

In general, discrepancy between the calculated and measured transport
properties of epitaxial magnetic multilayers can often be ascribed
to imperfect atomic structure at the interfaces. 
The correct treatment of structure defects on an \emph{ab initio}
level employs either supercell techniques 
\cite{r_2001_xkb, r_2002_dkb} or effective medium 
approaches \cite{r_2006_ctk} combined with a particular
microscopic model of the structure imperfection.  
In order to get a rough insight into the sensitivity of the TAMR to
the quality of interfaces, we adopted here a simplified approach.
We have simulated chemical disorder in the system by a finite
imaginary part $\varepsilon > 0$ of the energy arguments 
$z = E_{\rm F} \pm {\rm i} \varepsilon$ of the potential functions
in the TB-LMTO Kubo-Landauer formalism \cite{r_2000_kdb, r_2006_ctk}.
This modification was used only in a few selected atomic planes of
the whole system: in two neighbouring planes located at a single 
interface (Fe/As or As/Ag), at both interfaces, inside the GaAs
barrier, or inside the metallic electrodes.
The value of $\varepsilon = 5$~mRy was used in all cases.
The results for $n=27$ (the thickness corresponding to the maximum
TAMR effect in the perfect junctions) are collected in 
table~\ref{tabone}.
One can see that the disorder deep inside each part (Fe, GaAs, Ag)
of the junction has only a minor influence on the resulting TAMR.
However, the interface disorder reduces the TAMR value quite
significantly, which proves that at least a part of the difference
between the large theoretical values of the TAMR and the much
weaker observed effect is due to the interface roughness.
Moreover, both interfaces influence the TAMR effect to a similar
extent, see table~\ref{tabone}, which indicates that they are of
equal importance for the calculated trend of the TAMR 
(figure~\ref{f_thick}(b)).

\begin{figure}
\begin{center}
\rotatebox{270}{\includegraphics[width=0.35\textwidth]{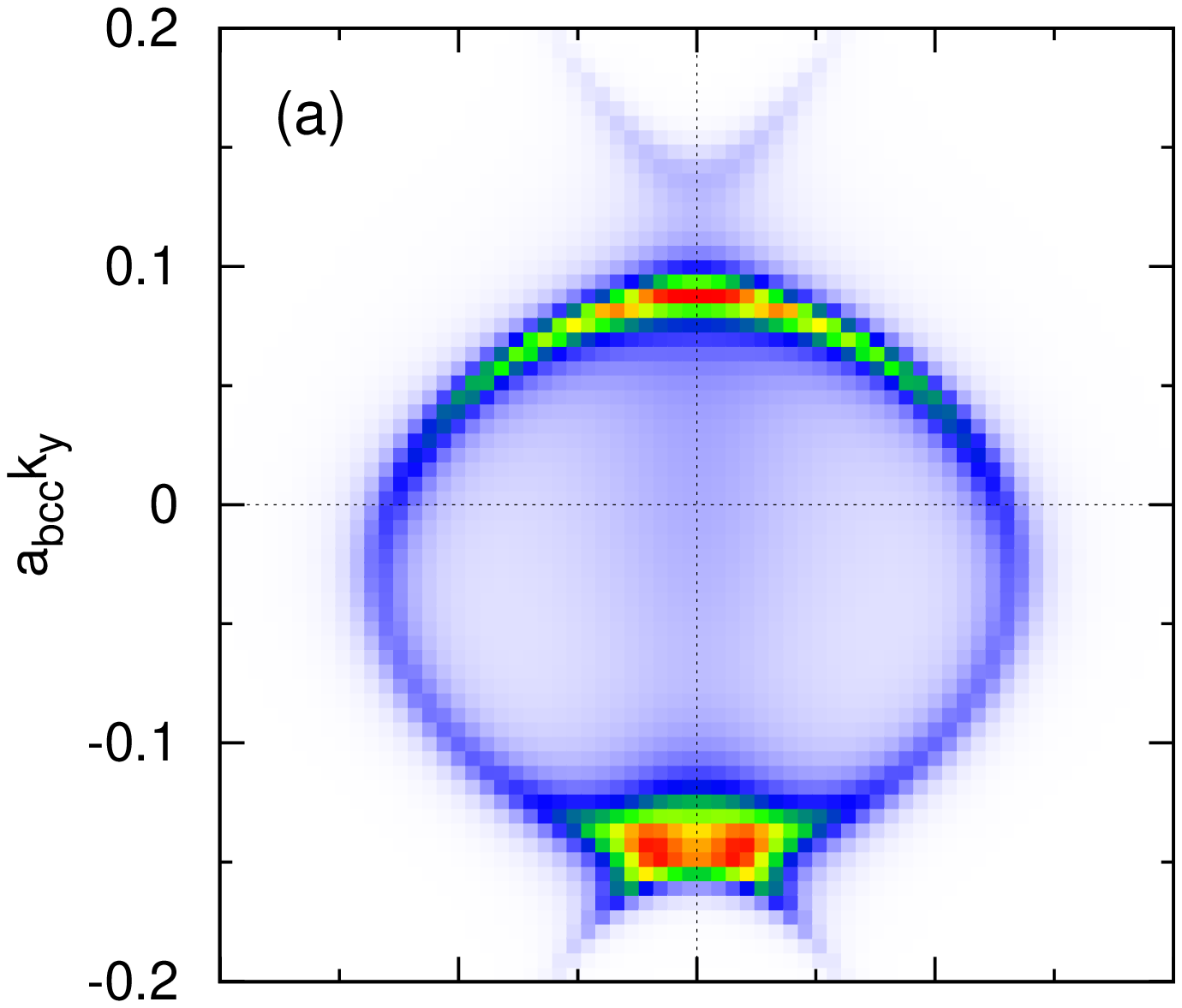}}
\rotatebox{270}{\includegraphics[width=0.35\textwidth]{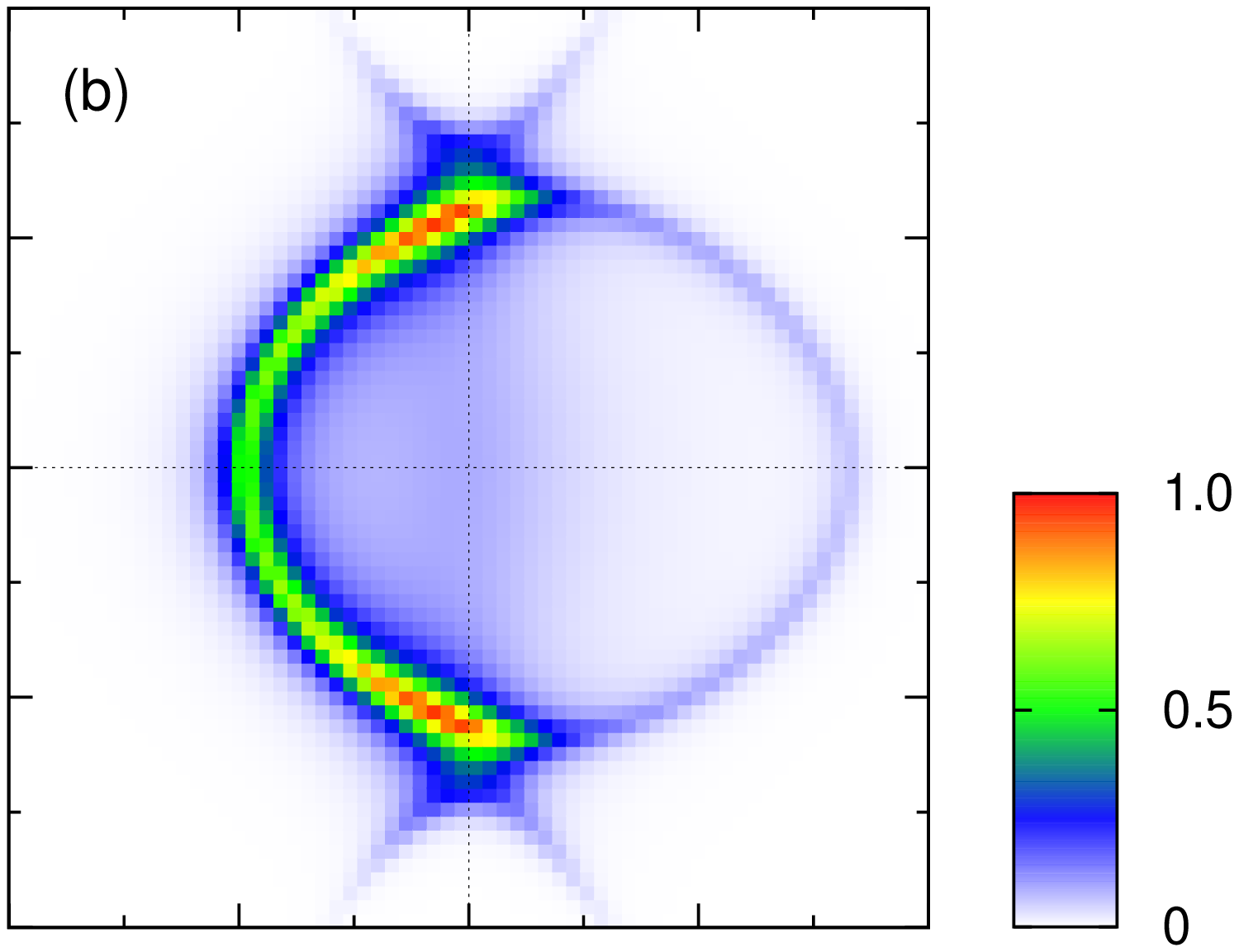}}

\rotatebox{270}{\includegraphics[width=0.35\textwidth]{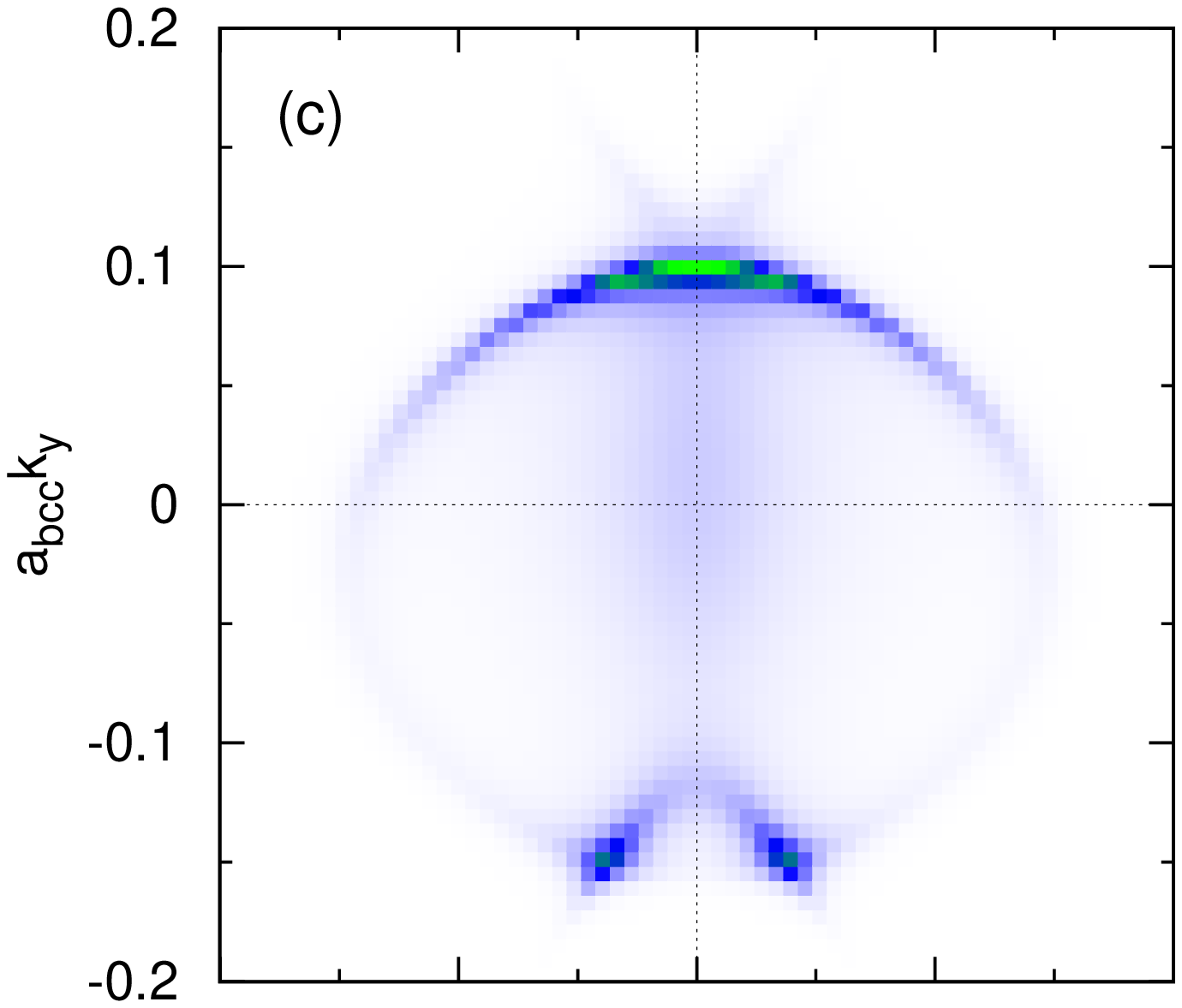}}
\rotatebox{270}{\includegraphics[width=0.35\textwidth]{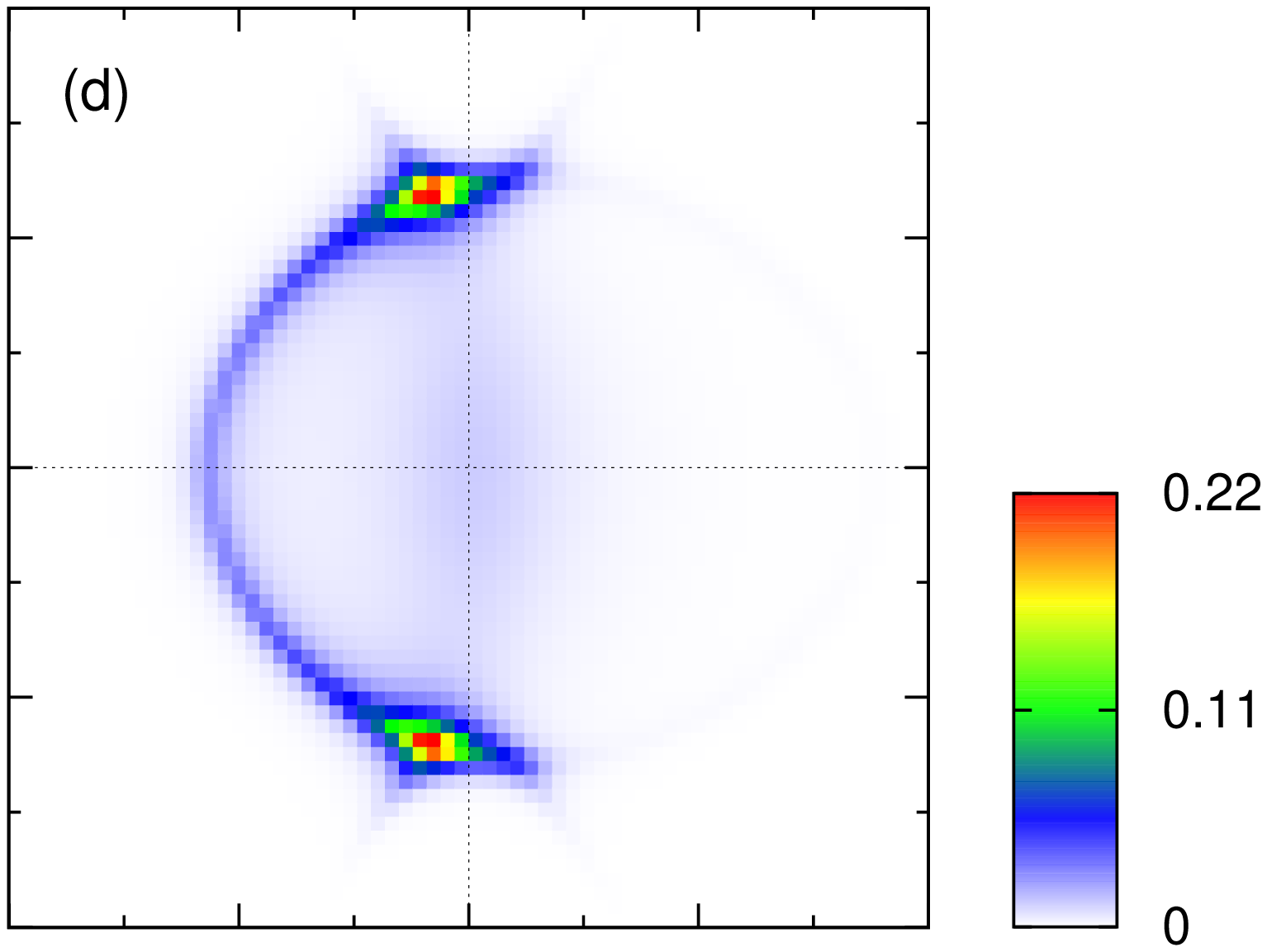}}

\rotatebox{270}{\includegraphics[width=0.40\textwidth]{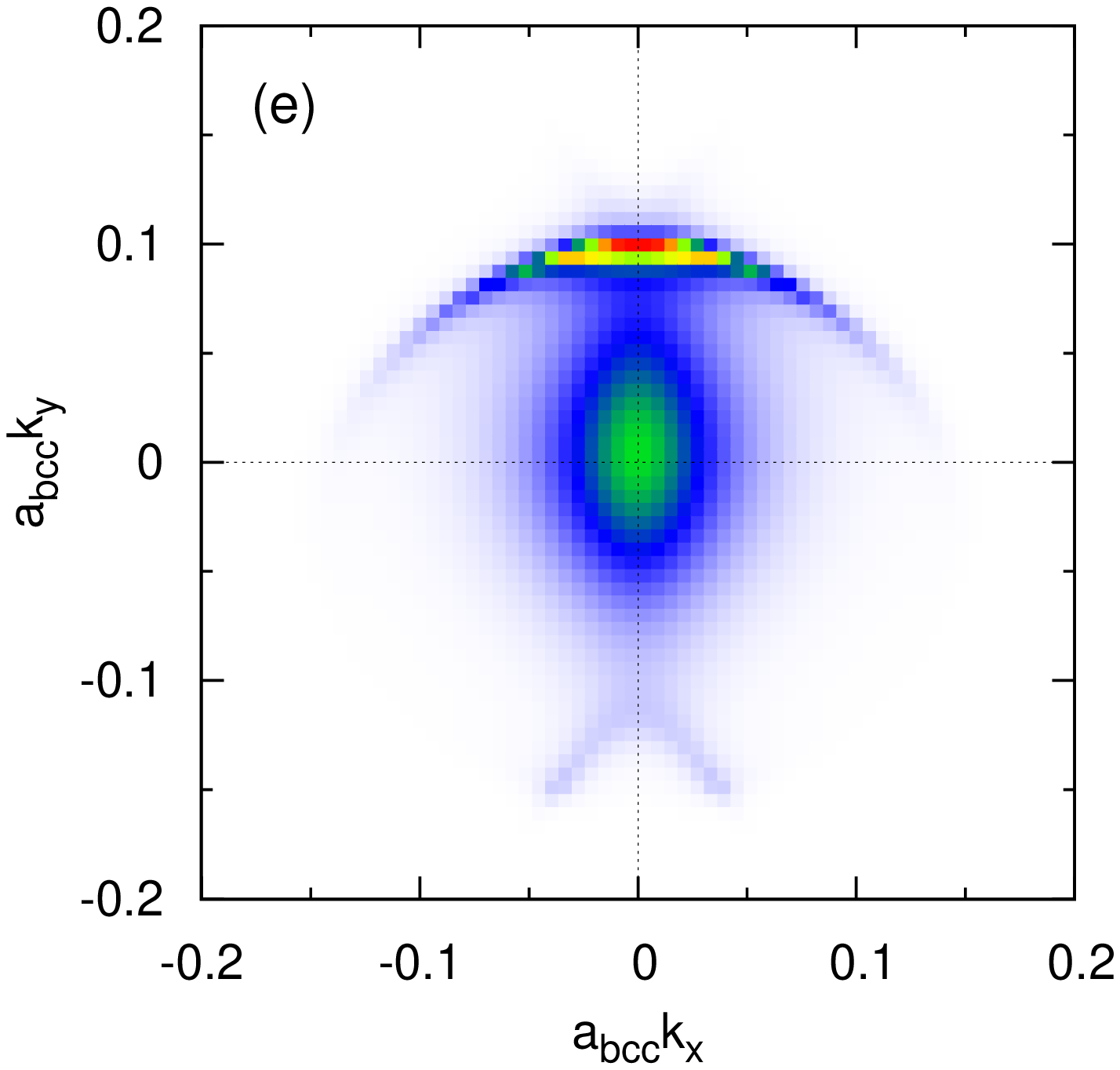}}
\rotatebox{270}{\includegraphics[width=0.40\textwidth]{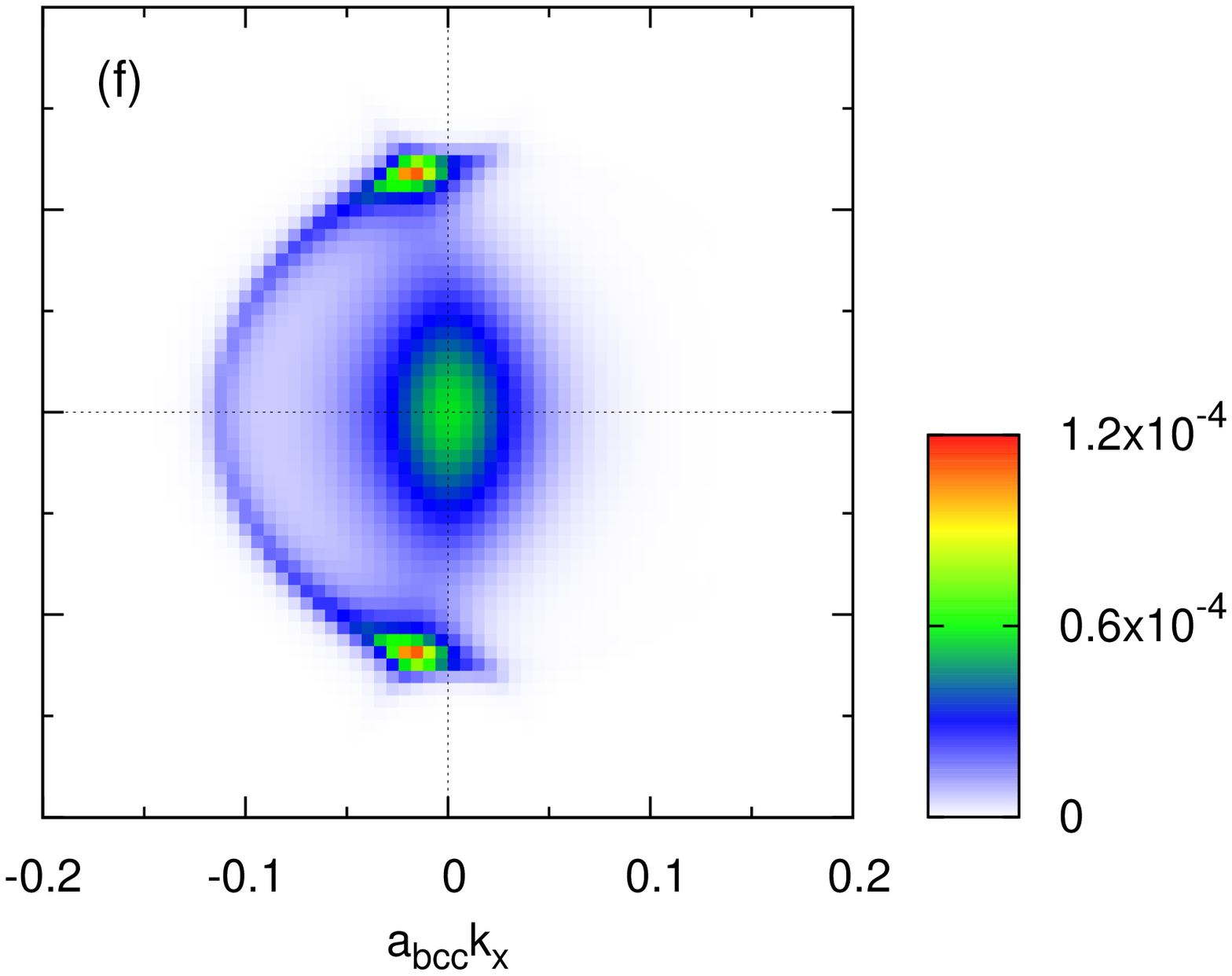}}
\end{center}
\caption{%
The $\bi{k}_\|$-resolved transmissions of the
Fe/As(GaAs)$_{n-1}$/Ag(001) systems at the Fermi energy
for $\theta = \pi/2$: 
(a) $n=19$, $\phi=0$, (b) $n=19$, $\phi=\pi/2$, 
(c) $n=27$, $\phi=0$, (d) $n=27$, $\phi=\pi/2$, 
(e) $n=47$, $\phi=0$, and (f) $n=47$, $\phi=\pi/2$. 
The coloured scales of the $T(\bi{k}_\|)$, shown on the right,
refer to both plots with the same $n$. 
\label{f_kres}}
\end{figure}

Further insight into the obtained results follows from
the $\bi{k}_\|$-resolved transmissions $T(\bi{k}_\|)$. 
Figure~\ref{f_kres} shows this quantity for three barrier
thicknesses, $n=19$, $n=27$, and $n=47$, and for the
magnetization directions along the $x$ and $y$ axis.
Only a small region around the $\bar{\Gamma}$ point is included
in the figure while the rest of the entire 2D BZ (defined by
$a_{\rm bcc} | k_{x,y} | \le \pi/ \sqrt{2} \approx 2.22$) is
unimportant for the tunnelling.
The maximum TAMR ($n=27$) corresponds to a few hot spots
in the $T(\bi{k}_\|)$ plots, see figure~\ref{f_kres}(c, d);
their positions and contributions to the total conductance are
strongly sensitive to the direction of the iron magnetic moment. 
The lower TAMR values for thinner barriers are due to non-negligible
contributions of bigger 2D regions to the total conductances, see
the case of $n=19$ in figure~\ref{f_kres}(a, b).
On the other hand, the lower TAMR values for very thick barriers
($n > 40$) are due to the same hot spots with non-zero
$\bi{k}_\|$-vectors as found for intermediate thicknesses 
($n \approx 27$) accompanied by another pronounced local maximum
of the $T(\bi{k}_\|)$ in the $\bar{\Gamma}$ point, 
see the case of $n=47$ in figure~\ref{f_kres}(e, f).
The contribution of the latter maximum to the total conductance
is little sensitive to the magnetization orientation which
explains the reduction of the TAMR effect.

The presence of the hot spots in the $\bi{k}_\|$-resolved
transmissions is undoubtedly an important factor contributing
to the high TAMR and to its non-monotonic dependence on the GaAs
thickness.
Similar hot spots appeared in various magnetic tunnel
junctions \cite{r_2002_wpz, r_2002_dmw, r_2005_bvt, r_2007_tbv}
and were shown to be a direct consequence of a hybridization of
two interface resonances across the barrier \cite{r_2002_wpz}.
In the present case, an interplay of these hybridization-induced
hot spots with the contribution of the $\bar{\Gamma}$ point
represents a new situation, relevant especially for large
thicknesses of barriers with a direct band gap, such as MgO or
GaAs \cite{r_2002_dmw}.

\section{Hybridized interface resonances in 
         a tight-binding model\label{s_hirtbm}}

In order to assess the role of the presence and hybridization
of the interface resonances on the TAMR, we have formulated
a simple TB model and investigated its properties.
The atoms are placed in positions of a simple cubic lattice
with the lattice parameter $a$; the atomic planes are the
(001) planes. 
The active part of the FM/I/NM junction comprises $N+2$ atomic
planes labelled by an index $p$, $p = 0, 1, \dots , N+1$.
The plane $p=0$ corresponds to a FM layer, the plane $p=N+1$
corresponds to a NM layer, and the tunnel barrier is represented
by the planes $p = 1, 2, \dots , N$.  
We assume a single orbital per site and spin and a spin-independent
nearest-neighbour hopping between the orbitals.
The hopping elements are different for pairs of atoms in neighbouring
atomic planes (hopping $t$) and inside the same atomic plane
(hopping $\tilde{t}$). 
The atomic energy levels of the tunnel barrier and of the NM layer
are spin-independent while those of the FM layer are exchange-split.
The direction of the exchange splitting is given by an in-plane
unit vector $\bi{n} = (\cos \phi, \sin \phi, 0)$.
The FM layer is also influenced by a Rashba-type SO splitting
derived from the canonical form $H^{\mathrm{SO}} \sim 
(\bi{p} \times \bsigma) \cdot \bnu$, where the
$\bi{p} = (p_x, p_y, p_z)$ denotes the momentum operator,
the $\bsigma = ( \sigma_x, \sigma_y, \sigma_z )$ are the Pauli
spin matrices and the unit vector $\bnu = (0, 0, 1)$ is normal
to the atomic planes.
We assume full 2D translational symmetry so that the Hamiltonian
of the system can be written after the 2D lattice Fourier
transformation as:
\begin{eqnarray}
H_{ps,p's'} (\bi{k}_\|) & = & \delta_{pp'} \left\{ h^{(p)}_{ss'} 
- 2 \tilde{t} [ \cos (ak_x) + \cos (ak_y) ] \delta_{ss'} \right\} 
\nonumber\\
 & & {} - \delta_{|p-p'|,1} \delta_{ss'} t
 + \delta_{p,0} \delta_{p',0} 
H^{\mathrm{SO}}_{ss'} (\bi{k}_\|) ,
\label{eq_tbham}
\end{eqnarray}
where the $s$ and $s'$ are spin indices, 
$s,s' = \uparrow, \downarrow$, which refer to the global (fixed)
spin quantization axis along the (001) direction.
The term $h^{(p)}_{ss'}$ comprises all on-site interactions in the
$p$-th atomic plane and the last term describes the Rashba-type SO
interaction in the FM layer ($p=0$). 
For the NM layer ($p=N+1$), the on-site term is given by 
$h^{(N+1)}_{ss'} = \epsilon_\mathrm{NM} \delta_{ss'}$, and
for the barrier layers ($p = 1, 2, \dots, N$), it is given
similarly as $h^{(p)}_{ss'} = \epsilon_\mathrm{B} \delta_{ss'}$, 
where the parameters $\epsilon_\mathrm{NM}$
and $\epsilon_\mathrm{B}$ denote, respectively, the energy levels
of the NM and barrier layers.
The explicit form of the FM on-site term ($p=0$) is given by
\begin{equation}
h^{(0)}_{\uparrow, \uparrow} = h^{(0)}_{\downarrow, \downarrow} 
 =  \frac{ \epsilon_\uparrow + \epsilon_\downarrow }{2} ,
\qquad
h^{(0)}_{\uparrow, \downarrow} = 
h^{(0) \ast}_{\downarrow, \uparrow } = 
\frac{ \epsilon_\uparrow - \epsilon_\downarrow }{2} 
\exp ( - \mathrm{i} \phi ) ,
\label{eq_h0}
\end{equation}
where the parameters $\epsilon_\uparrow$ and $\epsilon_\downarrow$
denote the exchange-split energy levels.
The last term in (\ref{eq_tbham}) is given by
\begin{eqnarray}
H^{\mathrm{SO}}_{\uparrow, \uparrow} (\bi{k}_\|) =  
H^{\mathrm{SO}}_{\downarrow, \downarrow} (\bi{k}_\|) & = & 0 ,
\nonumber\\
H^{\mathrm{SO}}_{\uparrow, \downarrow} (\bi{k}_\|) =  
H^{\mathrm{SO}}_{\downarrow, \uparrow} (\bi{k}_\|)^\ast & = & 
- \alpha [ \sin (ak_y) + \mathrm{i} \sin (ak_x) ] ,
\label{eq_hso}
\end{eqnarray}
where the parameter $\alpha$ scales the Rashba-like SO interaction.
The effect of the semiinfinite FM and NM leads has been simplified
by adding energy- and $\bi{k}_\|$-independent selfenergy 
operators to the on-site interactions of the NM and FM layers.
These (retarded) selfenergies are given by
\begin{eqnarray}
\Sigma^{(\mathrm{NM})}_{ss'} = - \frac{\mathrm{i}}{2} 
\gamma_\mathrm{NM} \delta_{ss'} ,
& \qquad &
\Sigma^{(\mathrm{FM})}_{\uparrow, \uparrow} = 
\Sigma^{(\mathrm{FM})}_{\downarrow, \downarrow} = 
- \frac{\mathrm{i}}{4} (\gamma_\uparrow + \gamma_\downarrow) ,
\nonumber\\
\Sigma^{(\mathrm{FM})}_{\uparrow, \downarrow} = 
- \frac{\mathrm{i}}{4} (\gamma_\uparrow - \gamma_\downarrow)
\exp( - \mathrm{i} \phi ) ,
& \qquad &
\Sigma^{(\mathrm{FM})}_{\downarrow, \uparrow} = 
- \frac{\mathrm{i}}{4} (\gamma_\uparrow - \gamma_\downarrow)
\exp( \mathrm{i} \phi ) ,
\label{eq_rse}
\end{eqnarray}
where the parameters $\gamma_\mathrm{NM}$, $\gamma_\uparrow$ and
$\gamma_\downarrow$ define the widths of the respective energy
levels (resonance widths in the local spin reference system).

\begin{figure}
\begin{center}
\includegraphics[width=0.50\textwidth]{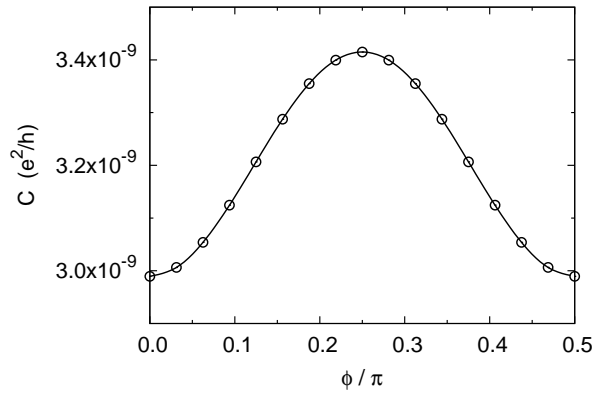}
\end{center}
\caption{%
Dependence of the conductance of the FM/I/NM model on the angle
$\phi$ for the barrier thickness $N = 30$ and for the case 2.
\label{f_angle_tb}}
\end{figure}

The simplicity of the model allows one to choose easily its
parameters in order to achieve the presence or absence of the
resonance at either interface and for each spin channel. 
Here we confine ourselves to the case of one spin channel 
(spin-$\downarrow$) out of the resonance and with no propagating
states in the FM lead; the latter condition is obtained by setting
$\gamma_\downarrow = 0$. 
We have considered four cases of the model. 
The first case, denoted as case 2, corresponds to the presence
of two resonances: one at the FM/I interface in the
spin-$\uparrow$ channel, the other at the I/NM interface.
The particular values of the model parameters are:
$t = 0.48$, $\tilde{t} = 0.03$, 
$\epsilon_\uparrow = 0.36$,  $\epsilon_\downarrow = -0.1$, 
$\epsilon_\mathrm{B} = 1.1$, $\epsilon_\mathrm{NM} = 0.4$,
$\gamma_\uparrow = 0.009$, $\gamma_\mathrm{NM} = 0.007$,
and $\alpha = 0.03$. 
Note that a small asymmetry has been intentionally introduced
in the parameters $\epsilon_\uparrow$/$\epsilon_\mathrm{NM}$ and
$\gamma_\uparrow$/$\gamma_\mathrm{NM}$ in order to simulate 
different properties of the FM and NM electrodes.
The Fermi energy is set to zero, $E_\mathrm{F} = 0$, which is
located slightly below the bottom of the spectrum of the tunnel
barrier, $V_\mathrm{B} = \epsilon_\mathrm{B} - 2t - 4\tilde{t} 
= 0.02$. 
The second case, denoted as case 1F, corresponds to the case
of one resonance, located at the FM/I interface for the
spin-$\uparrow$ channel; its parameters coincide with the case 2
apart from the value of $\epsilon_\mathrm{NM} = 0.6$.
The third case, denoted as case 1N, describes the situation with
one resonance, present at the I/NM interface. 
This case is obtained from the case 2 by setting the
value of $\epsilon_\uparrow = 0.6$. 
The last case, denoted as case 0, refers to the absence of any
resonance; its parameters are obtained from the case 2 by 
changing its two parameters, namely, 
$\epsilon_\uparrow = 0.6$ and $\epsilon_\mathrm{NM} = 0.64$.

\begin{figure}
\begin{center}
\includegraphics[width=0.80\textwidth]{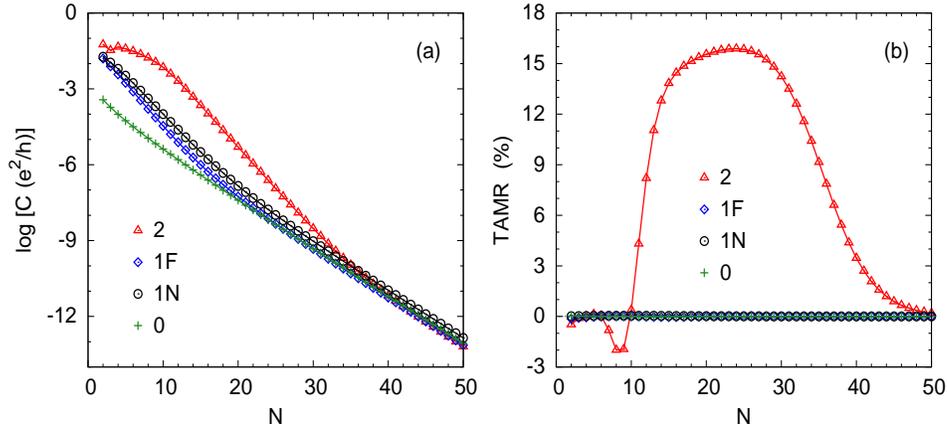}
\end{center}
\caption{%
Dependence of the transport properties of the FM/I/NM model on the
barrier thickness $N$ for the four cases (for details, see text):
(a) the conductance for $\phi = 0$, (b) the in-plane TAMR.
\label{f_thick_tb}}
\end{figure}

The angular dependence of the conductance $C(\phi)$ is plotted
in figure~\ref{f_angle_tb} for the case 2 and the barrier
thickness $N = 30$.
It is seen that the calculated dependence reflects the fourfold
symmetry of the system (point group $C_{\rm 4v}$).
The in-plane TAMR is thus defined as the ratio 
$[C(\pi/4) - C(0)]/C(0)$. 
The dependence of the conductances on the barrier thickness
for all four cases and for $\phi = 0$ is shown in 
figure~\ref{f_thick_tb}(a).
One can see clearly the effect of the resonances, pronounced 
especially for smaller thicknesses: 
the straight line for the case 0 is slightly modified by the
presence of a single resonance (cases 1F and 1N), whereas the
hybridization of two resonances is manifested by a strong
enhancement of the conductance (case 2). 
The corresponding thickness dependences of the TAMR are depicted
in figure~\ref{f_thick_tb}(b).
It is seen that a sizeable TAMR effect is obtained only for
the case 2 while the single resonances (on either side of the
barrier) lead essentially to negligible TAMR values
(cases 1F and 1N), similarly to the case 0. 
The hybridized interface resonances yield also a non-trivial
dependence of the TAMR on the barrier thickness: small
initial values for $N \leq 10$ are followed by a steep increase
to a broad maximum for $20 \leq N \leq 30$ which is replaced by
a final decrease for $N \geq 40$. 
This trend is qualitatively similar to that obtained for the
Fe/GaAs/Ag system, see figure~\ref{f_thick}(b).

\begin{figure}
\begin{center}
\rotatebox{270}{\includegraphics[width=0.35\textwidth]{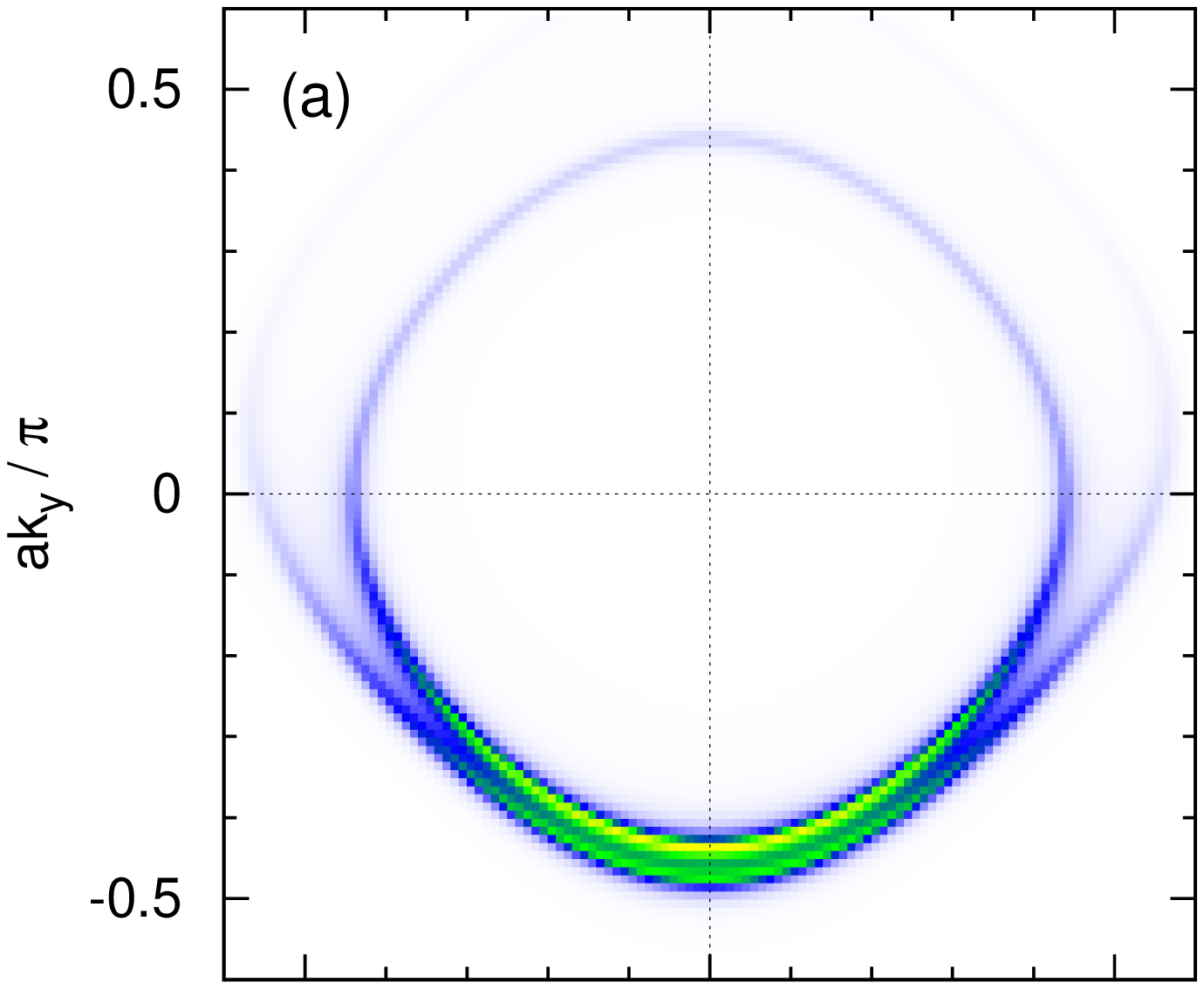}}
\rotatebox{270}{\includegraphics[width=0.35\textwidth]{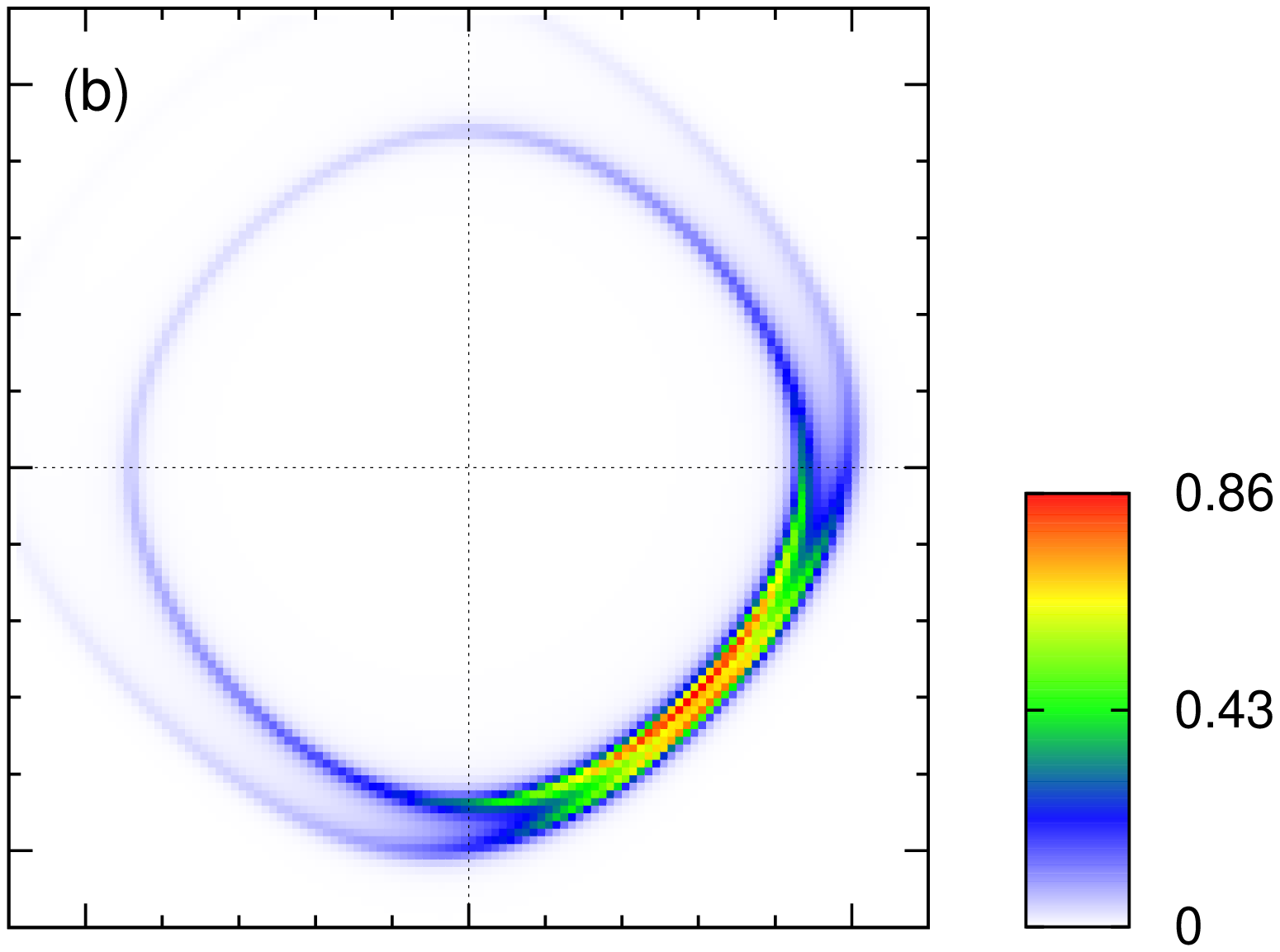}}

\rotatebox{270}{\includegraphics[width=0.35\textwidth]{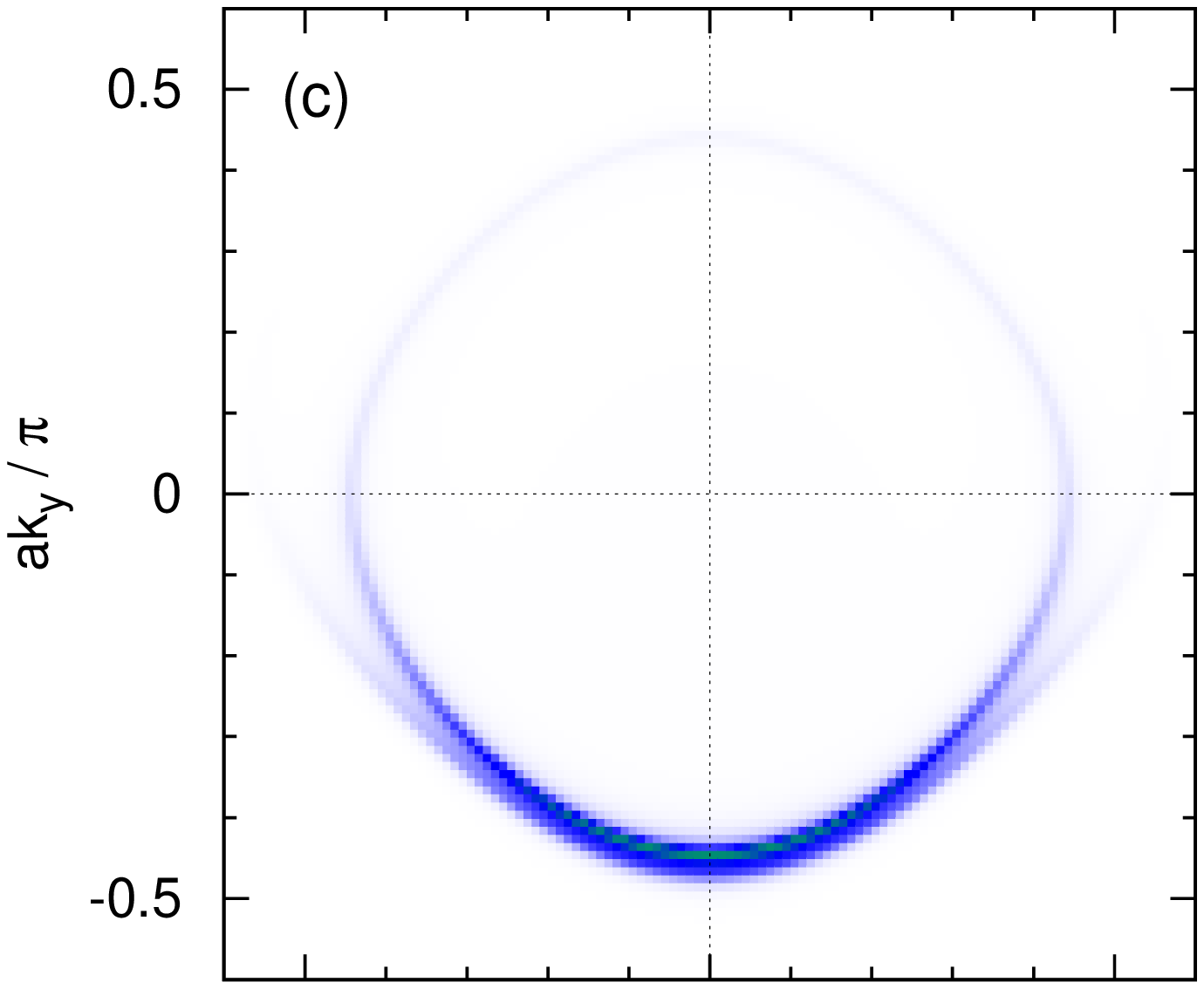}}
\rotatebox{270}{\includegraphics[width=0.35\textwidth]{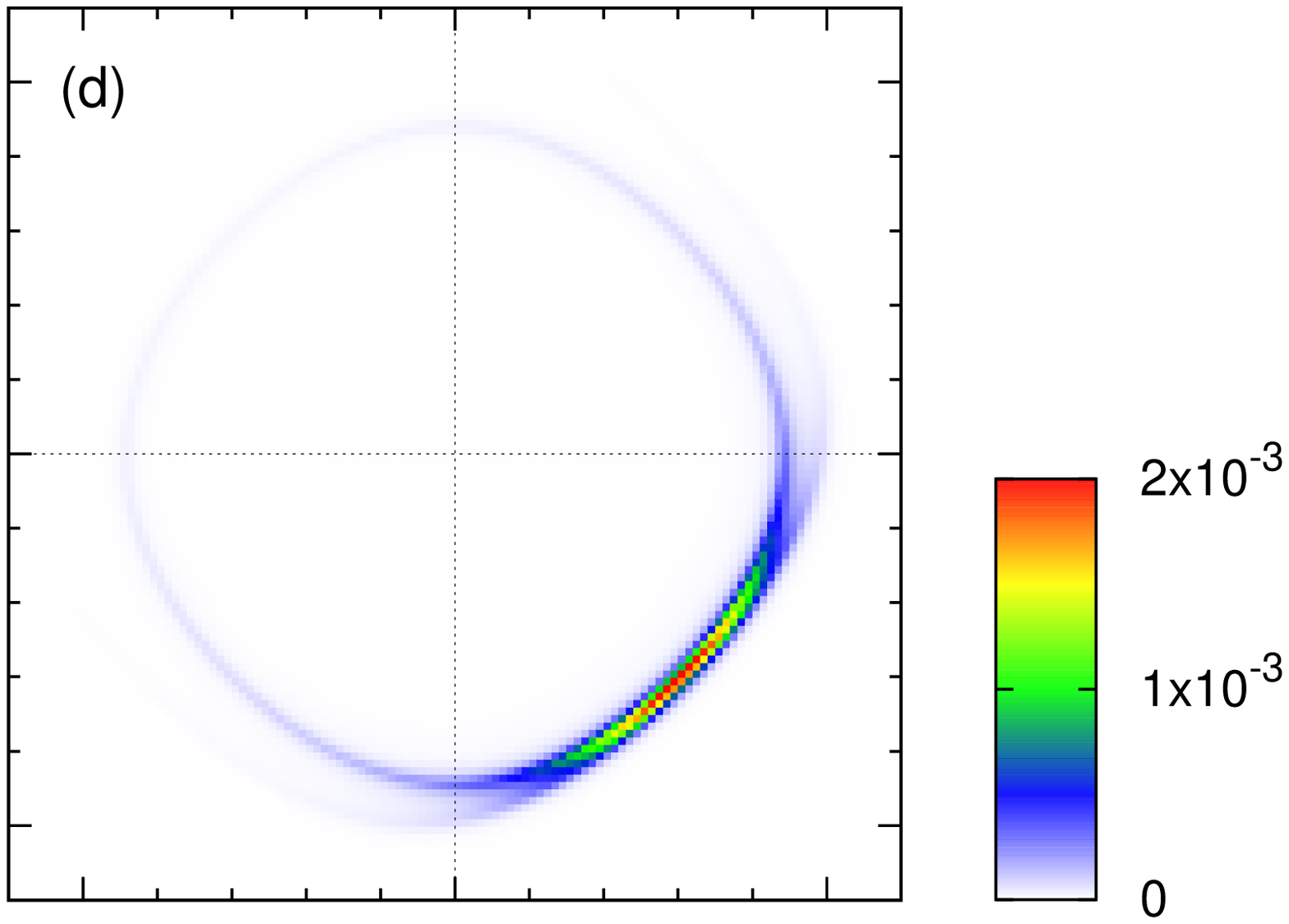}}

\rotatebox{270}{\includegraphics[width=0.40\textwidth]{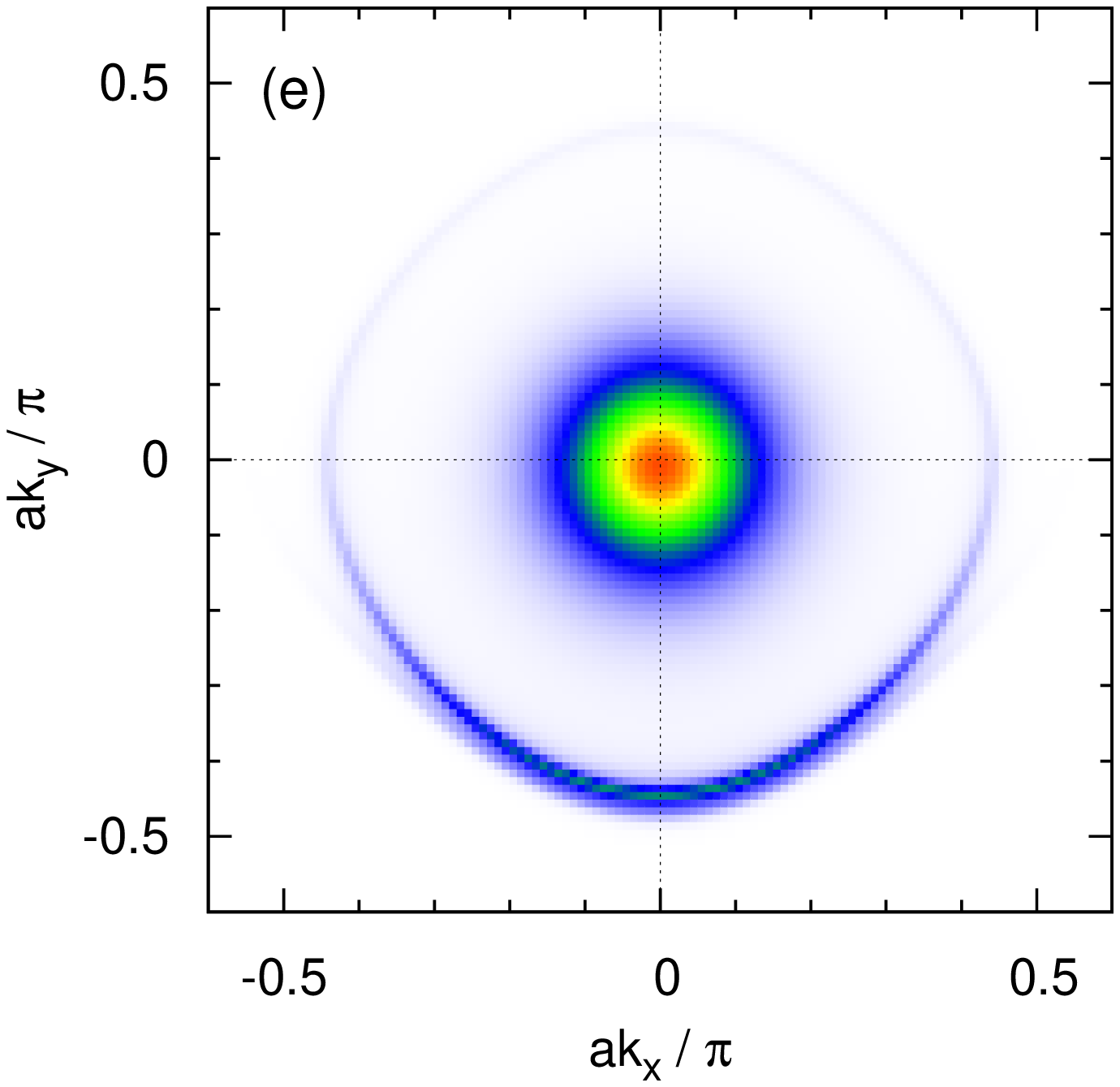}}
\rotatebox{270}{\includegraphics[width=0.40\textwidth]{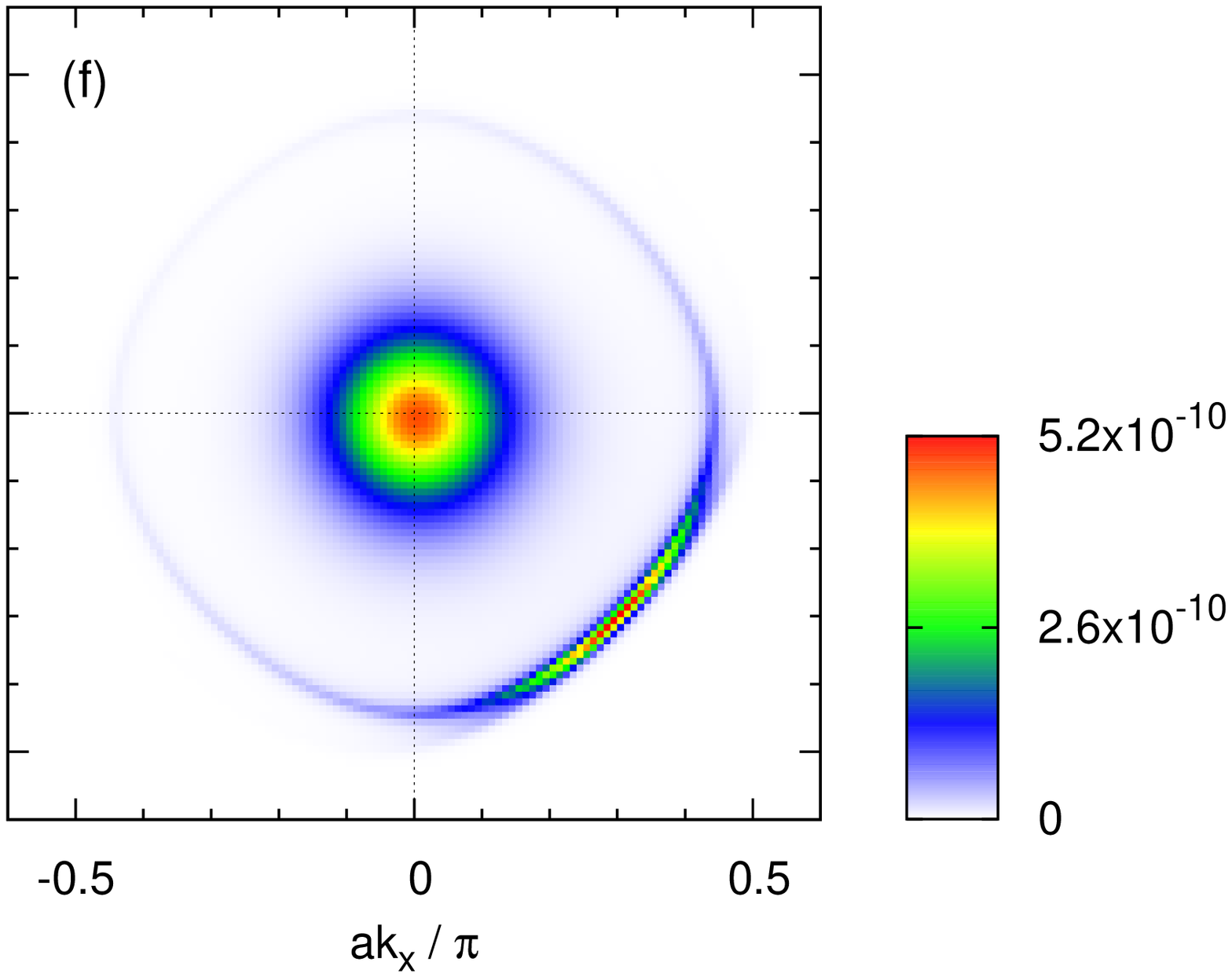}}
\end{center}
\caption{%
The $\bi{k}_\|$-resolved transmissions of the
FM/I/NM model in the case 2:
(a) $N=10$, $\phi=0$, (b) $N=10$, $\phi=\pi/4$, 
(c) $N=20$, $\phi=0$, (d) $N=20$, $\phi=\pi/4$, 
(e) $N=40$, $\phi=0$, and (f) $N=40$, $\phi=\pi/4$. 
The coloured scales of the $T(\bi{k}_\|)$, shown on the right,
refer to both plots with the same $N$. 
\label{f_kres_tb}}
\end{figure}

The different regimes of the thickness dependence of the TAMR
can be related to the corresponding $\bi{k}_\|$-resolved
transmissions shown in figure~\ref{f_kres_tb} for the case 2
and three values of $N$.
For $N=10$, the total conductances arise from contributions of
substantial parts of the whole 2D BZ which leads to a modest TAMR
effect for thin tunnelling barriers.
For $N=20$, the dominating contribution to the tunnelling is due to
a narrow region (a hot spot), the position of which depends on
the angle $\phi$. 
The sensitivity of this sharp single local maximum to the angle
$\phi$ gives rise to enhanced TAMR values for intermediate barrier
thicknesses. 
For $N=40$, the hot spots survive but are accompanied by a pronounced
peak in the very centre of the 2D BZ, which is reflected by reduced
TAMR values for very large insulator thicknesses $N$.
This reduction is a simple consequence of vanishing of the Rashba
term $H^{\mathrm{SO}}_{ss'} (\bi{k}_\|)$ in the limit of
$k_x \to 0$, $k_y \to 0$, see (\ref{eq_hso}).
The obtained changes in the $\bi{k}_\|$-resolved transmissions
are in close analogy to the first-principles results, which
corroborates the conclusions drawn in section~\ref{ss_cond}.

\section{Conclusions\label{s_conc}}

The calculated \emph{ab initio} results for the Fe/GaAs/Ag(001)
system with perfect epitaxial interfaces and the properties of a
simple tight-binding model of the FM/I/NM junctions demonstrate that
hybridized interface resonances can strongly influence the TAMR.
In particular, they can lead to sizable TAMR values, especially
for intermediate thicknesses of the tunnel barriers. 
The hybridized interface resonances can be thus added to the list
of existing origins of the TAMR: the anisotropic 
density of states of the FM electrode \cite{r_2004_grj},
the interference effects of the Rashba and Dresselhaus
contributions to the SO interaction \cite{r_2007_mms, r_2009_mf},
and the interface states at the FM/I interface \cite{r_2007_cbt}. 
This new mechanism -- if realized experimentally in a special
junction -- might also be employed to enhance the TAMR effect 
for applications in spintronics.
Further open problems related to the presented results, such as,
e.g., the effect of a finite bias, external magnetic fields or 
elevated temperatures, remain a task for future studies.

\ack
This work was supported financially by 
the Czech Science Foundation (Grant No.\ P204/11/1228).

\section*{References}


\end{document}